\documentclass[useAMS]{mn2e}

\usepackage{graphicx}
\newcommand{\teff}{\mbox{$T_{\rm eff}$}}
\newcommand{\logg}{\mbox{$\log g$}}
\newcommand{\vsini}{\mbox{$v \sin i$}}

\newcommand{\etal}{et\,al.}

\def\sqiglt{\hbox{\rlap{\lower.55ex \hbox {$\sim$}}
        \kern-.3em \raise.4ex \hbox{$<$}\,}}
\def\sqiggt{\hbox{\rlap{\lower.55ex \hbox {$\sim$}}
        \kern-.3em \raise.4ex \hbox{$>$}\,}}

\title[WASP-South hot Jupiters]{Transiting hot Jupiters from WASP-South, Euler and TRAPPIST: WASP-95b to WASP-101b}

\author[Hellier et al.]{Coel Hellier$^{1}$, 
D.R. Anderson$^{1}$, 
A. Collier Cameron$^{2}$, 
L. Delrez$^{3}$,
M. Gillon$^{3}$,\newauthor  
E. Jehin$^{3}$, 
M. Lendl$^{4}$, 
P.F.L. Maxted$^{1}$, 
F. Pepe$^{4}$, 
D. Pollacco$^{5}$, 
D. Queloz$^{4,6}$, \newauthor  
D. S\'egransan$^{4}$, 
B. Smalley$^{1}$, 
A.M.S. Smith$^{1,7}$, 
J. Southworth$^{1}$, \newauthor  
A.H.M.J. Triaud$^{4,8}$\thanks{Fellow of the Swiss National Science Foundation},  
S. Udry$^{4}$ \&\ 
R.G. West$^{5}$\\    
$^{1}$Astrophysics Group, Keele University, Staffordshire, ST5 5BG, UK\\
$^{2}$SUPA, School of Physics and Astronomy, University of St.\ Andrews, North Haugh,  Fife, KY16 9SS, UK\\
$^{3}$Institut d'Astrophysique et de G\'eophysique, Universit\'e de
Li\`ege, All\'ee du 6 Ao\^ut, 17, Bat. B5C, Li\`ege 1, Belgium\\
$^{4}$Observatoire astronomique de l'Universit\'e de Gen\`eve
51 ch. des Maillettes, 1290 Sauverny, Switzerland\\
$^{5}$Department of Physics, University of Warwick, Gibbet Hill Road, Coventry CV4 7AL, UK\\
$^{6}$Cavendish Laboratory, J J Thomson Avenue, Cambridge, CB3 0HE, UK\\
$^{7}$N.~Copernicus Astronomical Centre, Polish Academy of Sciences, Bartycka 18, 00-716 Warsaw, Poland\\
$^{8}$Department of Physics and Kavli Institute for Astrophysics \&\ Space Research, Massachusetts Institute of Technology,\\ Cambridge, MA 02139, USA\\
}

\begin{document}

\date{date}
\pagerange{range}

\maketitle

\begin{abstract}
We report the discovery of the transiting exoplanets WASP-95b, WASP-96b, WASP-97b, WASP-98b, WASP-99b, WASP-100b and WASP-101b.  All are hot Jupiters
with orbital periods in the range 2.1 to 5.7 d, masses of 0.5 to 2.8  M$_{\rm Jup}$, and radii of 1.1 to  1.4 R$_{\rm Jup}$.  The orbits of all the planets are compatible with zero eccentricity.  WASP-99b shows the shallowest transit yet found by WASP-South, at 0.4\%.   

The host stars are of spectral type F2 to G8. Five have metallicities of [Fe/H] from --0.03 to +0.23, while WASP-98 has a metallicity of --0.60, exceptionally low for a star with a transiting exoplanet.  Five of the host stars are brighter than  $V$ = 10.8, which significantly extends the number of bright transiting systems available for follow-up studies.  WASP-95 shows a possible rotational modulation at a period of 20.7 d.  

We discuss the completeness of WASP survey techniques by comparing to the HAT project.
 \end{abstract}

\begin{keywords}
planetary systems
\end{keywords}

\begin{table}
\caption{Observations\protect\rule[-1.5mm]{0mm}{2mm}}  
\begin{tabular}{lcr}
\hline 
Facility & Date &  \\ [0.5mm] \hline
\multicolumn{3}{l}{{\bf WASP-95:}}\\  
WASP-South & 2010 May--2011 Nov & 23\,200 points \\
Euler/CORALIE  & 2012 Jul--2013 May  &   14 radial velocities \\
TRAPPIST & 2012 Sep 14 & $z$ band \\
EulerCAM  & 2012 Sep 16 & Gunn r filter \\ 
TRAPPIST & 2012 Sep 27 & $z$ band \\
EulerCAM  & 2013 May 23 & Gunn r filter \\ 
EulerCAM  & 2013 Aug 01 & Gunn r filter \\ [0.5mm]
\multicolumn{3}{l}{\bf WASP-96:}\\  
WASP-South & 2010 Jun--2011 Dec & 13\,100 points \\
Euler/CORALIE  & 2011 Oct--2012 Oct  &   21 radial velocities \\
TRAPPIST & 2011 Nov 11 & Blue-block filter \\
EulerCAM  & 2012 Jul 01 & Gunn $r$ filter \\ 
TRAPPIST & 2012 Dec 11 & Blue-block filter \\
EulerCAM  & 2013 Jun 29 & Gunn $r$ filter \\ [0.5mm] 
\multicolumn{3}{l}{\bf WASP-97:}\\  
WASP-South & 2010 Jun--2012 Jan  & 23\,900 points \\
Euler/CORALIE  & 2012 Sep--2012 Nov  &   12 radial velocities \\
EulerCAM  & 2012 Nov 23 & Gunn $r$ filter \\ 
TRAPPIST & 2012 Nov 23 & $z$-band \\
EulerCAM  & 2013 Jul 11 & Gunn $r$ filter \\ 
EulerCAM  & 2013 Aug 09 & Gunn $r$ filter \\ 
TRAPPIST & 2013 Aug 09 & $I+z$ filter \\ [0.5mm]
\multicolumn{3}{l}{\bf WASP-98:}\\  
WASP-South & 2006 Aug--2012 Jan & 12\,700 points \\
Euler/CORALIE  & 2011 Nov--2012 Nov  &   15 radial velocities \\
EulerCAM  & 2012 Oct 28 & $I_{\rm C}$ band \\ 
TRAPPIST & 2012 Oct 28 & blue-block filter \\
EulerCAM  & 2012 Oct 31 & $I_{\rm C}$ band \\ 
TRAPPIST & 2012 Nov 03 & blue-block filter \\
TRAPPIST & 2013 Aug 29 & blue-block filter \\ 
EulerCAM  & 2013 Sep 07 & $I_{\rm C}$ band \\ [0.5mm]
\multicolumn{3}{l}{\bf WASP-99:}\\  
WASP-South & 2010 Jul--2012 Jan & 9000 points \\
Euler/CORALIE  & 2012 Feb--2013 Jan  &   20 radial velocities \\
EulerCAM  & 2012 Nov 21 & Gunn $r$ filter \\ [0.5mm] 
\multicolumn{3}{l}{\bf WASP-100:}\\ [0.5mm] 
WASP-South & 2010 Aug--2012 Jan & 13\,500 points \\
Euler/CORALIE  & 2012 Sep--2013 Mar  &   19 radial velocities \\
TRAPPIST & 2012 Dec 04 & $I+z$ filter \\ 
TRAPPIST & 2013 Jan 10 & $I+z$ filter \\ 
TRAPPIST & 2013 Sep 29 & $z$ filter \\ [0.5mm] 
\multicolumn{3}{l}{\bf WASP-101:}\\ [0.5mm] 
WASP-South & 2009 Jan--2012 Mar & 16\,100 points \\
Euler/CORALIE  & 2011 Jan--2013 Jun  &   21 radial velocities \\
TRAPPIST & 2012 Jan 11 & $z$ band \\
TRAPPIST & 2012 Jan 22 & $z$ band \\
TRAPPIST & 2012 Jan 29 & $z$ band \\
TRAPPIST & 2012 Mar 05 & $z$ band \\
EulerCAM  & 2012 Dec 06 & Gunn $r$ filter \\ 
TRAPPIST & 2013 Feb 23 & $z$ band \\ [0.5mm] \hline
\end{tabular} 
\end{table}

\section{Introduction}
The WASP-South survey has dominated the discovery of transiting hot-Jupiter exoplanets in the Southern hemisphere. WASP-South is well matched to the capabilities of the Euler/CORALIE spectrograph and the robotic  TRAPPIST telescope, with the combination of all three proving efficient for discovering transiting exoplanets in the range  $V$ = 9--13.

WASP-South has now been running nearly continuously for 7 years. Approximately 1000 candidates have been observed with Euler/CORALIE, while, since December 2010, TRAPPIST has observed 1000 lightcurves of WASP candidates and planets. Here we present new WASP-South planets which take WASP numbering above 100. 

Since WASP host stars are generally brighter than host stars of {\it Kepler\/} exoplanets, ongoing WASP-South discoveries are important for detailed study of exoplanets and will be prime targets for future missions such as {\it CHEOPS\/} and {\it JWST,} and proposed missions such as {\it EChO\/} and {\it FINESSE}. 

\section{Observations}
The observational and analysis techniques used here are the same as in recent WASP discovery papers (e.g.\ Hellier \etal\ 2012), and thus are described briefly.   For detailed accounts see the early papers including Pollacco \etal\ (2006), Collier-Cameron \etal\ (2007a) and Pollacco \etal\ (2007).

In outline, WASP-South surveys the visible sky each clear night using an array of 200mm f/1.8 lenses and a cadence of $\sim$\,10 mins. Transit searching of accumulated lightcurves leads to candidates that are passed to TRAPPIST (Jehin \etal\ 2011), a robotic 0.6-m photometric telescope, which can resolve blends and check that the candidate transits are planet-like, and to the 1.2-m Euler/CORALIE spectrograph, for radial-velocity observations.     About 1 in 12 candidates 
turns out to be a planet. Higher-quality transit lightcurves are then obtained with TRAPPIST and EulerCAM (Lendl \etal\ 2012).  A list of the observations reported here is given in Table~1 while the CORALIE radial velocities are listed in Table~A1.

\section{The host stars} 
For each star, the individual CORALIE spectra were co-added to produce a single
spectrum with typical S/N of $\sim$\,100:1 (though the fainter WASP-98, $V$ = 13,  had a S/N of only 40:1). Our methods for spectral analysis are described in Doyle \etal\ (2013).  The excitation balance of the Fe~{\sc i} lines was
used to determine the effective temperature (\teff), and spectral type was estimated from \teff\ using the table in Gray (2008). The surface gravity (\logg) was determined from the ionisation balance of Fe~{\sc i} and Fe~{\sc ii}. The Ca~{\sc i} line at 6439{\AA} and the Na~{\sc i} D lines were also used as \logg\ diagnostics. The metallicity was determined from equivalent width measurements of several unblended lines.

The projected stellar rotation velocity (\vsini) was determined by fitting the
profiles of several unblended Fe~{\sc i} lines. An instrumental FWHM of
0.11 $\pm$ 0.01~{\AA} was determined from the telluric lines around 6300\AA. 

For age estimates we use the lithium abundance and the Sestito \&\ Randlich (2005) calibration. We also give a gyrochronological age, from the measured \vsini\ and assuming that the star's spin is perpendicular to us, so that this is the true equatorial speed.  This is then combined with the stellar radius to give a rotational period, to compare with the results of Barnes (2007).   The results for each star are listed in the Tables 2 to 10.    We also list proper motions  from the UCAC4 catalogue  of Zacharias \etal\ (2013). The motions and metallicities are all compatible with the stars being local thin-disc stars.  

We searched the WASP photometry of each star for rotational
modulations by using a sine-wave fitting algorithm as described
by Maxted \etal\ (2011). We estimated the
significance of periodicities by subtracting the fitted transit lightcurve
and then repeatedly and randomly permuting the nights of observation.   We found a marginally significant modulation in WASP-95 (see Section 4.2) but not in any of the other stars (with 95\%-confidence upper limits being typically 1 mmag).

\section{System parameters}
The CORALIE radial-velocity measurements were combined with the WASP,
EulerCAM and TRAPPIST photometry in a simultaneous Markov-chain
Monte-Carlo (MCMC) analysis to find the system parameters.
For details of our methods see Collier Cameron \etal\ (2007b). 
The limb-darkening parameters are noted in each Table, and are
taken from the 4-parameter non-linear
law of Claret (2000).

For all of our planets the data are compatible with zero eccentricity 
and hence we imposed a circular orbit (see Anderson \etal\
2012 for the rationale for this).  The upper limits on the eccentricity range from 0.02 (for the $V$ = 9.5 WASP-99) to 0.27 (for the fainter, $V$ = 12.2, WASP-96). 

\clearpage

\begin{table}
\caption{System parameters for WASP-95.}  
\begin{tabular}{lc}
\multicolumn{2}{l}{1SWASP\,J222949.73--480011.0}\\
\multicolumn{2}{l}{2MASS\,22294972--4800111}\\
\multicolumn{2}{l}{RA\,=\,22$^{\rm h}$29$^{\rm m}$49.73$^{\rm s}$, 
Dec\,=\,--48$^{\circ}$00$^{'}$11.0$^{''}$ (J2000)}\\
\multicolumn{2}{l}{$V$ mag = 10.1}  \\ 
\multicolumn{2}{l}{Rotational modulation\ \ \ 2 mmag at $\sim$\, 20.7 d}\\
\multicolumn{2}{l}{pm (RA) 94.1\,$\pm$\,0.9 (Dec) --8.5\,$\pm$\,1.0 mas/yr}\\
\hline
\multicolumn{2}{l}{Stellar parameters from spectroscopic analysis.\rule[-1.5mm]{
0mm}{2mm}} \\ \hline 
Spectral type & G2 \\
$T_{\rm eff}$ (K) & 5830   $\pm$ 140  \\
$\log g$      & 4.36 $\pm$ 0.07 \\
$v\,\sin I$ (km\,s$^{-1}$)     & 3.1 $\pm$ 0.6 \\
{[Fe/H]}   &   +0.14 $\pm$ 0.16 \\
log A(Li)  &   $<$0.2  \\
Age (Lithium) [Gy]  & $>$3     \\
Age (Gyro) [Gy]     &$2.4^{+1.7}_{-1.0}$ \\ \hline 
\multicolumn{2}{l}{Parameters from MCMC analysis.\rule[-1.5mm]{0mm}{3mm}} \\
\hline 
$P$ (d) & 2.1846730   $\pm$ 0.0000014 \\
$T_{\rm c}$ (HJD)\,(UTC) & 245\,6338.45851 $\pm$ 0.00024\\ 
$T_{\rm 14}$ (d) & 0.116 $\pm$ 0.001\\ 
$T_{\rm 12}=T_{\rm 34}$ (d) & 0.011 $\pm$ 0.001 \\
$\Delta F=R_{\rm P}^{2}$/R$_{*}^{2}$ & 0.0105 $\pm$ 0.0003\\ 
$b$ & 0.19  $^{+0.21}_{-0.13}$  \\
$i$ ($^\circ$) & 88.4 $^{+1.2}_{-2.1}$\\
$K_{\rm 1}$ (km s$^{-1}$) & 0.1757 $\pm$ 0.0017\\ 
$\gamma$ (km s$^{-1}$) & 6.2845 $\pm$ 0.0004\\ 
$e$ & 0 (adopted) ($<$\,0.04 at 3$\sigma$) \\ 
$M_{\rm *}$ (M$_{\rm \odot}$) & 1.11 $\pm$ 0.09\\ 
$R_{\rm *}$ (R$_{\rm \odot}$) & 1.13  $^{+0.08}_{-0.04}$\\
$\log g_{*}$ (cgs) & 4.38 $^{+0.02}_{-0.04}$ \\
$\rho_{\rm *}$ ($\rho_{\rm \odot}$) &0.78  $^{+0.04}_{-0.13}$ \\
$T_{\rm eff}$ (K) & 5630 $\pm$ 130\\
$M_{\rm P}$ (M$_{\rm Jup}$) & 1.13 $^{+0.1}_{-0.04}$\\
$R_{\rm P}$ (R$_{\rm Jup}$) & 1.21 $\pm$ 0.06\\
$\log g_{\rm P}$ (cgs) & 3.34 $^{+0.02}_{-0.07}$\\
$\rho_{\rm P}$ ($\rho_{\rm J}$) & 0.85 $^{+0..07}_{-0.2}$\\
$a$ (AU)  & 0.03416 $\pm$ 0.00083\\
$T_{\rm P, A=0}$ (K) & 1570 $\pm$ 50\\ [0.5mm] \hline 
\multicolumn{2}{l}{Errors are 1$\sigma$; Limb-darkening coefficients were:}\\
\multicolumn{2}{l}{(Euler $r$) a1 =    0.701, a2 = --0.490, a3 =  1.077, 
a4 = --0.516}\\ 
\multicolumn{2}{l}{(Trap $z$) a1 =    0.780, a2 = --0.718, a3 =  1.082, 
a4 = --0.487}\\ \hline
\end{tabular} 
\end{table}

\begin{figure}
\hspace*{-5mm}\includegraphics[width=10cm]{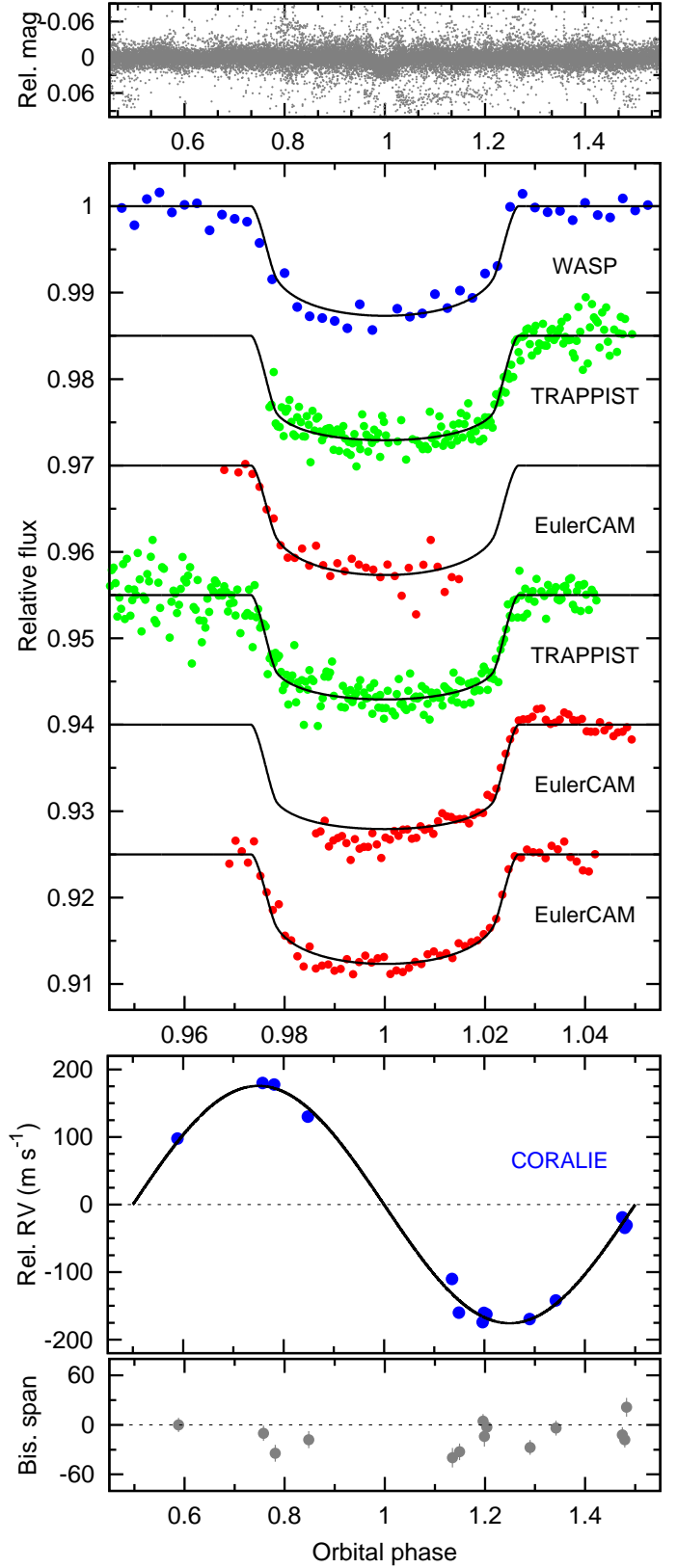}\\ [-2mm]
\caption{WASP-95b discovery data: (Top) The WASP data folded on the 
transit period. (Second panel) The binned WASP data with (offset) the
follow-up transit lightcurves (ordered from the top as 
in Table~1) together with the 
fitted MCMC model.  (Third) The CORALIE radial
velocities with the fitted model.
(Lowest) The bisector spans; the absence of any correlation with
radial velocity is a check against transit mimics.}
\end{figure}

\clearpage 

\begin{table}
\caption{System parameters for WASP-96.}  
\begin{tabular}{lc}
\multicolumn{2}{l}{1SWASP\,J000411.14--472138.2}\\
\multicolumn{2}{l}{2MASS\,00041112--4721382}\\
\multicolumn{2}{l}{RA\,=\,00$^{\rm h}$04$^{\rm m}$11.14$^{\rm s}$, 
Dec\,=\,--47$^{\circ}$21$^{'}$38.2$^{''}$ (J2000)}\\
\multicolumn{2}{l}{$V$ mag = 12.2}  \\ 
\multicolumn{2}{l}{Rotational modulation\ \ \ $<$\,1 mmag (95\%)}\\
\multicolumn{2}{l}{pm (RA) 23.1\,$\pm$\,1.0 (Dec) 3.7\,$\pm$\,1.0 mas/yr}\\
\hline
\multicolumn{2}{l}{Stellar parameters from spectroscopic analysis.\rule[-1.5mm]{
0mm}{2mm}} \\ \hline 
Spectral type & G8 \\
$T_{\rm eff}$ (K) & 5500     $\pm$ 150  \\
$\log g$      & 4.25 $\pm$ 0.15 \\
$v\,\sin I$ (km\,s$^{-1}$)     & 1.5 $\pm$ 1.3 \\
{[Fe/H]}   &   +0.14 $\pm$ 0.19 \\
log A(Li)  &    1.48 $\pm$ 0.15   \\
Age (Lithium) [Gy]  & 2\,$\sim$\,5     \\
Age (Gyro) [Gy]     & $8^{+26}_{-8}$ \\ \hline 
\multicolumn{2}{l}{Parameters from MCMC analysis.\rule[-1.5mm]{0mm}{3mm}} \\
\hline 
$P$ (d) & 3.4252602   $\pm$ 0.0000027 \\
$T_{\rm c}$ (HJD)\,(UTC) & 245\,6258.0621 $\pm$ 0.0002\\ 
$T_{\rm 14}$ (d) & 0.1011 $\pm$ 0.0011\\ 
$T_{\rm 12}=T_{\rm 34}$ (d) & 0.0200 $\pm$ 0.0014 \\
$\Delta F=R_{\rm P}^{2}$/R$_{*}^{2}$ & 0.0138 $\pm$ 0.0003\\ 
$b$ & 0.710  $\pm$ 0.019  \\
$i$ ($^\circ$) & 85.6 $\pm$ 0.2\\
$K_{\rm 1}$ (km s$^{-1}$) & 0.062 $\pm$ 0.004\\ 
$\gamma$ (km s$^{-1}$) & --1.23300 $\pm$ 0.00009\\ 
$e$ & 0 (adopted) ($<$\,0.27 at 3$\sigma$) \\ 
$M_{\rm *}$ (M$_{\rm \odot}$) & 1.06 $\pm$ 0.09\\ 
$R_{\rm *}$ (R$_{\rm \odot}$) & 1.05  $\pm$ 0.05\\
$\log g_{*}$ (cgs) & 4.42 $\pm$ 0.02 \\
$\rho_{\rm *}$ ($\rho_{\rm \odot}$) & 0.922 $\pm$ 0.073\\
$T_{\rm eff}$ (K) & 5540 $\pm$ 140\\
$M_{\rm P}$ (M$_{\rm Jup}$) & 0.48 $\pm$ 0.03\\
$R_{\rm P}$ (R$_{\rm Jup}$) & 1.20 $\pm$ 0.06\\
$\log g_{\rm P}$ (cgs) & 2.88 $\pm$ 0.04\\
$\rho_{\rm P}$ ($\rho_{\rm J}$) & 0.28 $\pm$ 0.04\\
$a$ (AU)  & 0.0453 $\pm$ 0.0013\\
$T_{\rm P, A=0}$ (K) & 1285 $\pm$ 40\\ [0.5mm] \hline 
\multicolumn{2}{l}{Errors are 1$\sigma$; Limb-darkening coefficients were:}\\
\multicolumn{2}{l}{(Euler $r$) a1 =    0.722, a2 = --0.581, a3 =  1.203, 
a4 = --0.561}\\ 
\multicolumn{2}{l}{(Trap BB) a1 =    0.722, a2 = --0.581, a3 =  1.203, 
a4 = --0.561}\\ \hline
\end{tabular} 
\end{table} 

\begin{figure}
\hspace*{-5mm}\includegraphics[width=10cm]{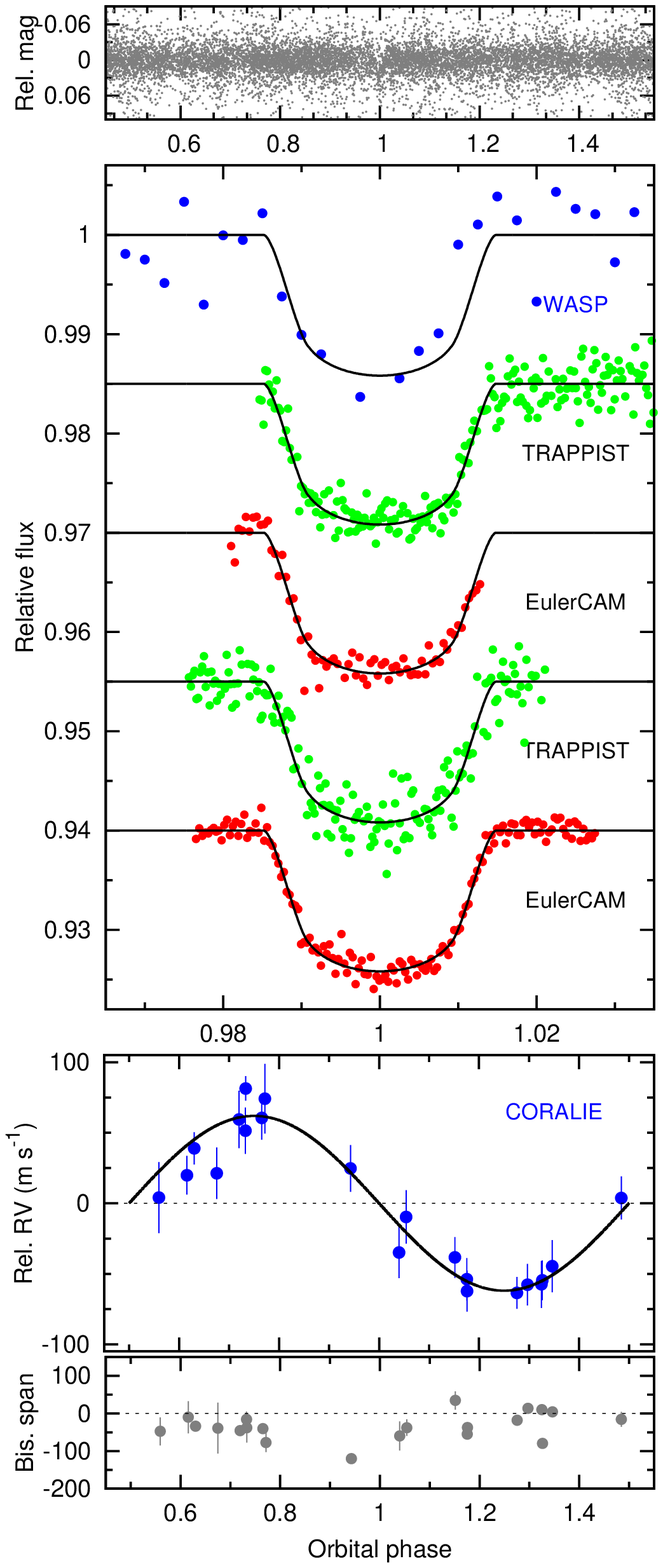}\\ [-2mm]
\caption{WASP-96b discovery data (as in Fig.~1).}
\end{figure}

\clearpage

\begin{table}
\caption{System parameters for WASP-97.}  
\begin{tabular}{lc}
\multicolumn{2}{l}{1SWASP\,J013825.04--554619.4}\\
\multicolumn{2}{l}{2MASS\,01382504--5546194}\\
\multicolumn{2}{l}{RA\,=\,01$^{\rm h}$38$^{\rm m}$25.04$^{\rm s}$, 
Dec\,=\,--55$^{\circ}$46$^{'}$19.4$^{''}$ (J2000)}\\
\multicolumn{2}{l}{$V$ mag = 10.6}  \\ 
\multicolumn{2}{l}{Rotational modulation\ \ \ $<$\,1 mmag (95\%)}\\
\multicolumn{2}{l}{pm (RA) 94.6\,$\pm$\,1.1 (Dec) 20.8\,$\pm$\,1.0 mas/yr}\\
\hline
\multicolumn{2}{l}{Stellar parameters from spectroscopic analysis.\rule[-1.5mm]{
0mm}{2mm}} \\ \hline 
Spectral type & G5 \\
$T_{\rm eff}$ (K)  & 5670     $\pm$ 110  \\
$\log g$      & 4.45 $\pm$ 0.08 \\
$v\,\sin I$ (km\,s$^{-1}$)     & 1.1 $\pm$ 0.5 \\
{[Fe/H]}   &   +0.23 $\pm$ 0.11 \\
log A(Li)  &   $<$ 0.85   \\
Age (Lithium) [Gy]  & $>$ 5     \\
Age (Gyro) [Gy]     &  11.9$^{+16.0}_{-8.3}$ \\ \hline 
\multicolumn{2}{l}{Parameters from MCMC analysis.\rule[-1.5mm]{0mm}{3mm}} \\
\hline 
$P$ (d) &    2.072760   $\pm$ 0.000001 \\
$T_{\rm c}$ (HJD)\,(UTC) & 245\,6438.18683 $\pm$ 0.00018\\ 
$T_{\rm 14}$ (d) & 0.1076 $\pm$ 0.0008\\ 
$T_{\rm 12}=T_{\rm 34}$ (d) &0.011  $\pm$ 0.001 \\
$\Delta F=R_{\rm P}^{2}$/R$_{*}^{2}$ & 0.0119 $\pm$ 0.0002\\ 
$b$ & 0.23 $^{+0.11}_{-0.15}$\\
$i$ ($^\circ$) & 88.0 $^{+1.3}_{-1.0}$\\
$K_{\rm 1}$ (km s$^{-1}$) & 0.1945  $\pm$ 0.0023\\ 
$\gamma$ (km s$^{-1}$) & 6.80877  $\pm$ 0.00029\\ 
$e$ & 0 (adopted) ($<$\,0.05 at 3$\sigma$) \\ 
$M_{\rm *}$ (M$_{\rm \odot}$) &  1.12 $\pm$ 0.06\\ 
$R_{\rm *}$ (R$_{\rm \odot}$) & 1.06  $\pm$ 0.04\\
$\log g_{*}$ (cgs) & 4.43 $\pm$ 0.03 \\
$\rho_{\rm *}$ ($\rho_{\rm \odot}$) &0.93  $\pm$ 0.09\\
$T_{\rm eff}$ (K) & 5640 $\pm$ 100\\
$M_{\rm P}$ (M$_{\rm Jup}$) & 1.32 $\pm$ 0.05\\
$R_{\rm P}$ (R$_{\rm Jup}$) & 1.13 $\pm$ 0.06\\
$\log g_{\rm P}$ (cgs) & 3.37 $\pm$ 0.04\\
$\rho_{\rm P}$ ($\rho_{\rm J}$) & 0.91 $\pm$ 0.11\\
$a$ (AU)  & 0.03303 $\pm$ 0.00056\\
$T_{\rm P, A=0}$ (K) & 1555 $\pm$ 40\\ [0.5mm] \hline 
\multicolumn{2}{l}{Errors are 1$\sigma$; Limb-darkening coefficients were:}\\
\multicolumn{2}{l}{(Euler $r$) a1 =    0.700, a2 = --0.489, a3 =  1.077, 
a4 = --0.516}\\ 
\multicolumn{2}{l}{(Trap $Iz$) a1 =    0.778, a2 = --0.713, a3 =  1.077, 
a4 = --0.484}\\ \hline
\end{tabular} 
\end{table}

\begin{figure}
\hspace*{-5mm}\includegraphics[width=10cm]{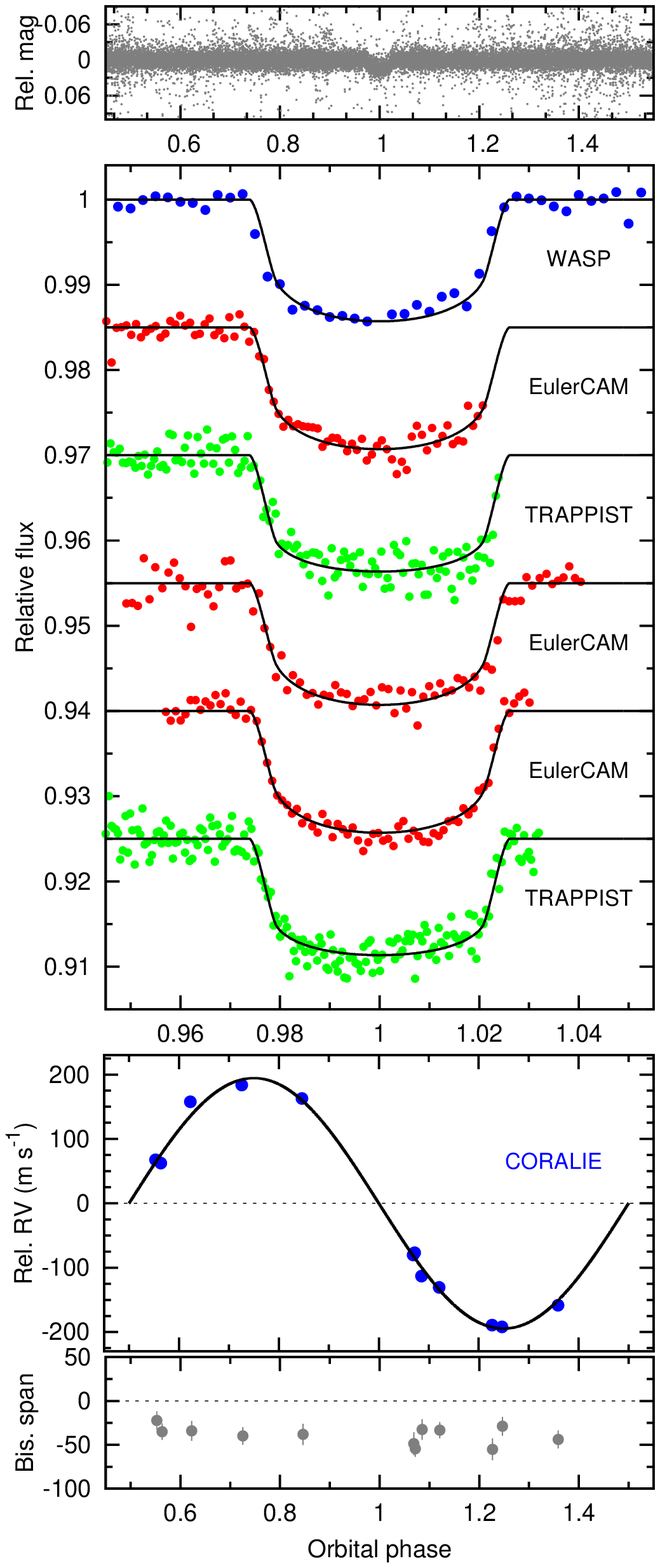}\\ [-2mm]
\caption{WASP-97b discovery data (as in Fig.~1).}
\end{figure}

\clearpage

\begin{table}
\caption{System parameters for WASP-98.}  
\begin{tabular}{lc}
\multicolumn{2}{l}{1SWASP\,J035342.90--341941.7}\\
\multicolumn{2}{l}{2MASS\,03534291--3419414}\\
\multicolumn{2}{l}{RA\,=\,03$^{\rm h}$53$^{\rm m}$42.90$^{\rm s}$, 
Dec\,=\,--34$^{\circ}$19$^{'}$41.7$^{''}$ (J2000)}\\
\multicolumn{2}{l}{$V$ mag = 13.0}  \\ 
\multicolumn{2}{l}{Rotational modulation\ \ \ $<$\,2 mmag (95\%)}\\
\multicolumn{2}{l}{pm (RA) 32.0\,$\pm$\,1.1 (Dec) --13.4\,$\pm$\,1.1 mas/yr}\\
\hline
\multicolumn{2}{l}{Stellar parameters from spectroscopic analysis.\rule[-1.5mm]{
0mm}{2mm}} \\ \hline 
Spectral type & G7 \\
$T_{\rm eff}$ (K) & 5550     $\pm$ 140  \\
$\log g$      & 4.40 $\pm$ 0.15 \\
$v\,\sin I$ (km\,s$^{-1}$)     &  $<$ 0.5   \\
{[Fe/H]}   &   $-$0.60 $\pm$ 0.19 \\
log A(Li)  &   $<$ 0.91   \\
Age (Lithium) [Gy]  & $>$ 3     \\
Age (Gyro) [Gy]     & $>$ 8  \\ \hline 
\multicolumn{2}{l}{Parameters from MCMC analysis.\rule[-1.5mm]{0mm}{3mm}} \\
\hline 
$P$ (d) &    2.9626400   $\pm$ 0.0000013 \\
$T_{\rm c}$ (HJD)\,(UTC) & 245\,6333.3913 $\pm$ 0.0001\\ 
$T_{\rm 14}$ (d) & 0.0795 $\pm$ 0.0005\\ 
$T_{\rm 12}=T_{\rm 34}$ (d) & 0.0202  $\pm$ 0.0006 \\
$\Delta F=R_{\rm P}^{2}$/R$_{*}^{2}$ &0.02570  $\pm$ 0.00025\\ 
$b$ & 0.71 $\pm$ 0.01 \\
$i$ ($^\circ$) & 86.3 $\pm$ 0.1 \\
$K_{\rm 1}$ (km s$^{-1}$) & 0.15  $\pm$ 0.01\\ 
$\gamma$ (km s$^{-1}$) & --38.2882  $\pm$ 0.0003\\ 
$e$ & 0 (adopted) ($<$\,0.24 at 3$\sigma$) \\ 
$M_{\rm *}$ (M$_{\rm \odot}$) & 0.69  $\pm$ 0.06\\ 
$R_{\rm *}$ (R$_{\rm \odot}$) & 0.70  $\pm$ 0.02\\
$\log g_{*}$ (cgs) & 4.583 $\pm$ 0.014 \\
$\rho_{\rm *}$ ($\rho_{\rm \odot}$) & 1.99  $\pm$ 0.07\\
$T_{\rm eff}$ (K) & 5525 $\pm$ 130\\
$M_{\rm P}$ (M$_{\rm Jup}$) & 0.83 $\pm$ 0.07\\
$R_{\rm P}$ (R$_{\rm Jup}$) & 1.10 $\pm$ 0.04\\
$\log g_{\rm P}$ (cgs) & 3.20 $\pm$ 0.03\\
$\rho_{\rm P}$ ($\rho_{\rm J}$) & 0.63 $\pm$ 0.06\\
$a$ (AU)  & 0.036 $\pm$ 0.001\\
$T_{\rm P, A=0}$ (K) & 1180 $\pm$ 30\\ [0.5mm] \hline 
\multicolumn{2}{l}{Errors are 1$\sigma$; Limb-darkening coefficients were:}\\
\multicolumn{2}{l}{(Euler $I$) a1 =    0.558, a2 = --0.157, a3 = 0.576, 
a4 = --0.325}\\ 
\multicolumn{2}{l}{(Trap BB) a1 =  0.486, a2 = 0.090, a3 =  0.444, 
a4 = --0.294}\\ \hline
\end{tabular} 
\end{table}

\begin{figure}
\hspace*{-5mm}\includegraphics[width=10cm]{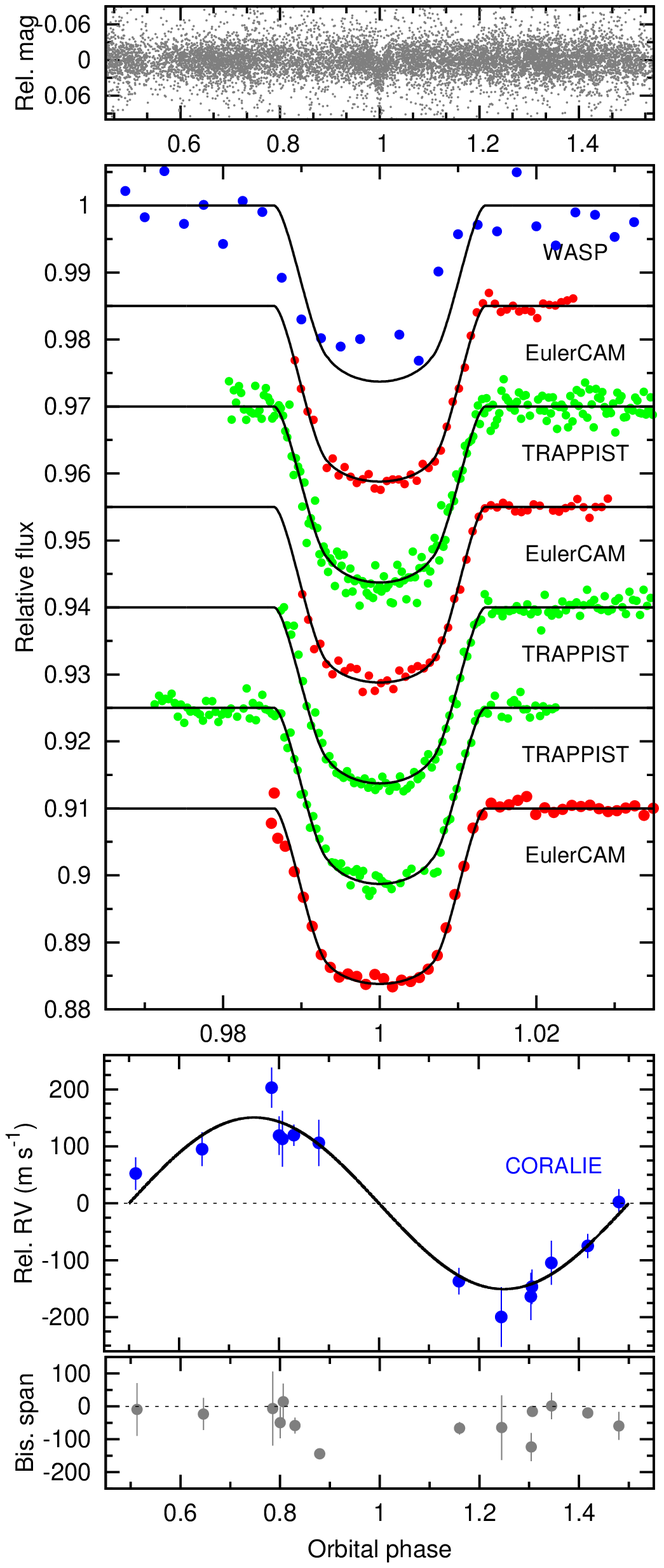}\\ [-2mm]
\caption{WASP-98b discovery data (as in Fig.~1).}
\end{figure}

\clearpage

\begin{table}
\caption{System parameters for WASP-99.}  
\begin{tabular}{lc}
\multicolumn{2}{l}{1SWASP\,J023935.44--500028.8}\\
\multicolumn{2}{l}{2MASS\,02393544--5000288}\\
\multicolumn{2}{l}{RA\,=\,02$^{\rm h}$39$^{\rm m}$35.44$^{\rm s}$, 
Dec\,=\,--50$^{\circ}$00$^{'}$28.8$^{''}$ (J2000)}\\
\multicolumn{2}{l}{$V$ mag = 9.5}  \\ 
\multicolumn{2}{l}{Rotational modulation\ \ \ $<$\,2 mmag (95\%)}\\
\multicolumn{2}{l}{pm (RA) --2.7\,$\pm$\,1.1 (Dec) --38.1\,$\pm$\,0.9 mas/yr}\\
\hline
\multicolumn{2}{l}{Stellar parameters from spectroscopic analysis.\rule[-1.5mm]{
0mm}{2mm}} \\ \hline 
Spectral type & F8 \\
$T_{\rm eff}$ (K) & 6150  $\pm$ 100  \\
$\log g$      &4.3  $\pm$ 0.1 \\
$v\,\sin I$ (km\,s$^{-1}$)     &    6.8 $\pm$ 0.5   \\
{[Fe/H]}   &    +0.21 $\pm$ 0.15 \\
log A(Li)  &  2.52 $\pm$ 0.08    \\
Age (Lithium) [Gy]  & 1\,$\sim$\,3       \\
Age (Gyro) [Gy]     & $1.4^{+1.1}_{-0.6}$   \\ \hline 
\multicolumn{2}{l}{Parameters from MCMC analysis.\rule[-1.5mm]{0mm}{3mm}} \\
\hline 
$P$ (d) &    5.75251   $\pm$ 0.00004 \\
$T_{\rm c}$ (HJD)\,(UTC) & 245\,6224.9824 $\pm$ 0.0014\\ 
$T_{\rm 14}$ (d) & 0.219 $\pm$ 0.003\\ 
$T_{\rm 12}=T_{\rm 34}$ (d) & 0.0137  $^{+0.0017}_{-0.0006}$ \\
$\Delta F=R_{\rm P}^{2}$/R$_{*}^{2}$ & 0.0041  $\pm$ 0.0002\\ 
$b$ & 0.18 $\pm$ 0.17 \\
$i$ ($^\circ$) & 88.8 $\pm$ 1.1 \\
$K_{\rm 1}$ (km s$^{-1}$) & 0.2422  $\pm$ 0.0017\\ 
$\gamma$ (km s$^{-1}$) & 24.9610  $\pm$ 0.0002\\ 
$e$ & 0 (adopted) ($<$\,0.02 at 3$\sigma$) \\ 
$M_{\rm *}$ (M$_{\rm \odot}$) &  1.48 $\pm$ 0.10\\ 
$R_{\rm *}$ (R$_{\rm \odot}$) &1.76   $^{+0.11}_{-0.06}$\\
$\log g_{*}$ (cgs) & 4.12 $^{+0.02}_{-0.04}$ \\
$\rho_{\rm *}$ ($\rho_{\rm \odot}$) & 0.27 $^{+0.02}_{-0.04}$\\
$T_{\rm eff}$ (K) & 6180 $\pm$ 100\\
$M_{\rm P}$ (M$_{\rm Jup}$) & 2.78 $\pm$ 0.13\\
$R_{\rm P}$ (R$_{\rm Jup}$) & 1.10 $^{+0.08}_{-0.05}$\\
$\log g_{\rm P}$ (cgs) & 3.72 $^{+0.03}_{-0.06}$\\
$\rho_{\rm P}$ ($\rho_{\rm J}$) & 2.1 $\pm$ 0.3\\
$a$ (AU)  & 0.0717 $\pm$ 0.0016\\
$T_{\rm P, A=0}$ (K) & 1480 $\pm$ 40\\ [0.5mm] \hline 
\multicolumn{2}{l}{Errors are 1$\sigma$; Limb-darkening coefficients were:}\\
\multicolumn{2}{l}{(Euler $r$) a1 =    0.590, a2 = 0.036, a3 =  0.306, 
a4 = --0.205}\\ \hline
\end{tabular} 
\end{table}

\begin{figure}
\hspace*{-5mm}\includegraphics[width=10cm]{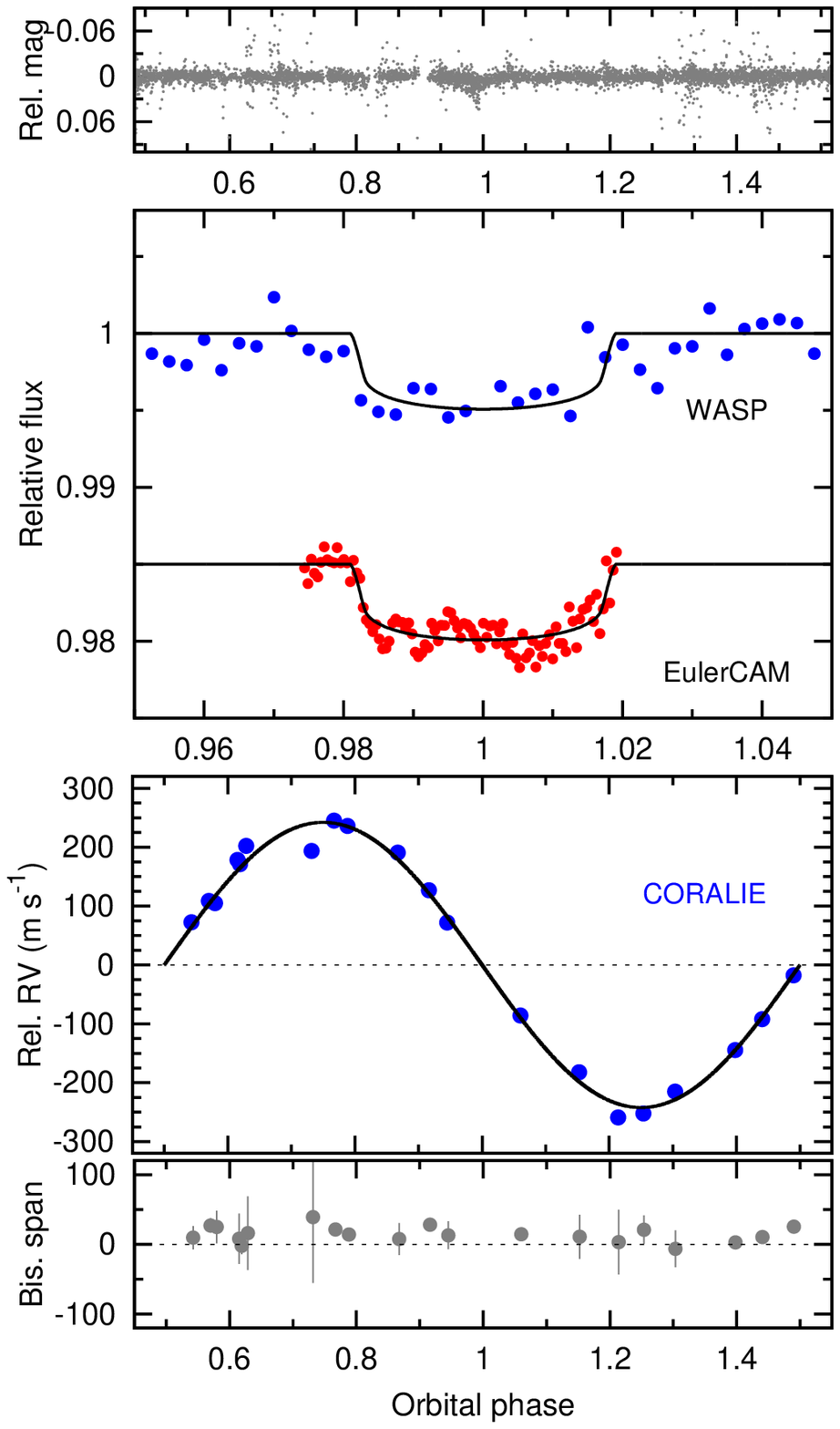}\\ [-2mm]
\caption{WASP-99b discovery data (as in Fig.~1).}
\end{figure}

\clearpage

\begin{table}
\caption{System parameters for WASP-100.}  
\begin{tabular}{lc}
\multicolumn{2}{l}{1SWASP\,J043550.32--640137.3}\\
\multicolumn{2}{l}{2MASS\,04355033--6401373}\\
\multicolumn{2}{l}{RA\,=\,04$^{\rm h}$35$^{\rm m}$50.32$^{\rm s}$, 
Dec\,=\,--64$^{\circ}$01$^{'}$37.3$^{''}$ (J2000)}\\
\multicolumn{2}{l}{$V$ mag = 10.8}  \\ 
\multicolumn{2}{l}{Rotational modulation\ \ \ $<$\,1 mmag (95\%)}\\
\multicolumn{2}{l}{pm (RA) 11.9\,$\pm$\,1.0 (Dec) --2.2\,$\pm$\,2.3 mas/yr}\\
\hline
\multicolumn{2}{l}{Stellar parameters from spectroscopic analysis.\rule[-1.5mm]{
0mm}{2mm}} \\ \hline 
Spectral type & F2 \\
$T_{\rm eff}$ (K) & 6900   $\pm$ 120  \\
$\log g$      &4.35  $\pm$ 0.17 \\
$v\,\sin I$ (km\,s$^{-1}$)     &    12.8 $\pm$ 0.8    \\
{[Fe/H]}   &   $-$0.03 $\pm$ 0.10   \\
log A(Li)  &  $<$ 1.80      \\
Age (Lithium) [Gy]  &  too hot      \\
Age (Gyro) [Gy]     &  too hot   \\ \hline 
\multicolumn{2}{l}{Parameters from MCMC analysis.\rule[-1.5mm]{0mm}{3mm}} \\
\hline 
$P$ (d) &  2.849375     $\pm$ 0.000008 \\
$T_{\rm c}$ (HJD)\,(UTC) & 245\,6272.3395 $\pm$ 0.0009\\ 
$T_{\rm 14}$ (d) & 0.160 $\pm$ 0.005\\ 
$T_{\rm 12}=T_{\rm 34}$ (d) & 0.021 $\pm$ 0.005 \\
$\Delta F=R_{\rm P}^{2}$/R$_{*}^{2}$ & 0.0076  $\pm$ 0.0005\\ 
$b$ & 0.64 $^{+0.08}_{-0.16}$ \\
$i$ ($^\circ$) & 82.6 $^{+2.6}_{-1.7}$ \\
$K_{\rm 1}$ (km s$^{-1}$) &  0.213 $\pm$ 0.008\\ 
$\gamma$ (km s$^{-1}$) &  29.9650 $\pm$ 0.0002\\ 
$e$ & 0 (adopted) ($<$\,0.10 at 3$\sigma$) \\ 
$M_{\rm *}$ (M$_{\rm \odot}$) &  1.57 $\pm$ 0.10\\ 
$R_{\rm *}$ (R$_{\rm \odot}$) & 2.0   $\pm$ 0.3\\
$\log g_{*}$ (cgs) & 4.04 $\pm$ 0.11 \\
$\rho_{\rm *}$ ($\rho_{\rm \odot}$) & 0.20 $^{+0.10}_{-0.05}$ \\
$T_{\rm eff}$ (K) & 6900 $\pm$ 120\\
$M_{\rm P}$ (M$_{\rm Jup}$) & 2.03 $\pm$ 0.12\\
$R_{\rm P}$ (R$_{\rm Jup}$) & 1.69 $\pm$ 0.29\\
$\log g_{\rm P}$ (cgs) & 3.21 $\pm$ 0.15 \\
$\rho_{\rm P}$ ($\rho_{\rm J}$) & 0.4 $\pm$ 0.2\\
$a$ (AU)  & 0.0457 $\pm$ 0.0010\\
$T_{\rm P, A=0}$ (K) & 2190 $\pm$ 140\\ [0.5mm] \hline 
\multicolumn{2}{l}{Errors are 1$\sigma$; Limb-darkening coefficients were:}\\
\multicolumn{2}{l}{(Trap $Iz$) a1 =  0.542, a2 = 0.086, a3 =  0.001, 
a4 = --0.055}\\ \hline
\end{tabular} 
\end{table}

\begin{figure}
\hspace*{-5mm}\includegraphics[width=10cm]{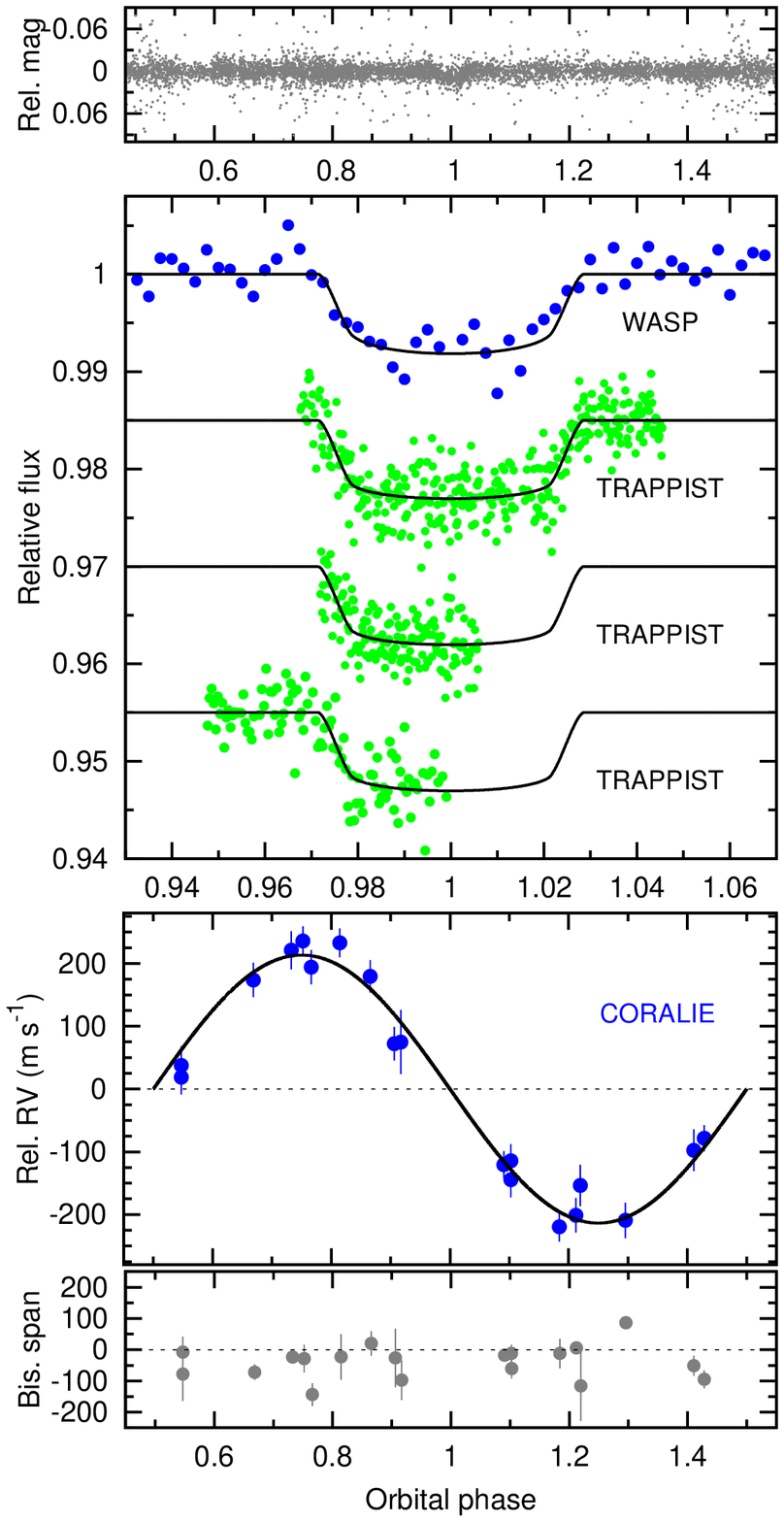}\\ [-2mm]
\caption{WASP-100b discovery data (as in Fig.~1).}
\end{figure}

\clearpage

\begin{table}
\caption{System parameters for WASP-101.}  
\begin{tabular}{lc}
\multicolumn{2}{l}{1SWASP\,J063324.26--232910.2}\\
\multicolumn{2}{l}{2MASS\,06332426--2329103}\\
\multicolumn{2}{l}{RA\,=\,06$^{\rm h}$33$^{\rm m}24.26$$^{\rm s}$, 
Dec\,=\,--23$^{\circ}$29$^{'}$10.2$^{''}$ (J2000)}\\
\multicolumn{2}{l}{$V$ mag = 10.3}  \\ 
\multicolumn{2}{l}{Rotational modulation\ \ \ $<$\,1 mmag (95\%)}\\
\multicolumn{2}{l}{pm (RA) --2.3\,$\pm$\,0.9 (Dec) 23.3\,$\pm$\,1.2 mas/yr}\\
\hline
\multicolumn{2}{l}{Stellar parameters from spectroscopic analysis.\rule[-1.5mm]{
0mm}{2mm}} \\ \hline 
Spectral type & F6 \\
$T_{\rm eff}$ (K)  & 6380  $\pm$ 120  \\
$\log g$      & 4.31 $\pm$ 0.08    \\
$v\,\sin I$ (km\,s$^{-1}$)     &    12.4 $\pm$ 0.5      \\
{[Fe/H]}   &   +0.20 $\pm$ 0.12     \\
log A(Li)  &    2.80 $\pm$ 0.09      \\
Age (Lithium) [Gy]  &    0.5\,$\sim$\,2          \\
Age (Gyro) [Gy]     &   $0.9^{+1.3}_{-0.4}$   \\ \hline  
\multicolumn{2}{l}{Parameters from MCMC analysis.\rule[-1.5mm]{0mm}{3mm}} \\
\hline 
$P$ (d) &    3.585722   $\pm$ 0.000004 \\
$T_{\rm c}$ (HJD)\,(UTC) & 245\,6164.6934 $\pm$ 0.0002\\ 
$T_{\rm 14}$ (d) & 0.113 $\pm$ 0.001\\ 
$T_{\rm 12}=T_{\rm 34}$ (d) & 0.023  $\pm$ 0.001  \\
$\Delta F=R_{\rm P}^{2}$/R$_{*}^{2}$ &  0.0126  $\pm$ 0.0002\\ 
$b$ & 0.736 $\pm$ 0.013 \\
$i$ ($^\circ$) & 85.0 $\pm$ 0.2 \\
$K_{\rm 1}$ (km s$^{-1}$) &  0.054 $\pm$ 0.004\\ 
$\gamma$ (km s$^{-1}$) &  42.6373 $\pm$ 0.0006\\ 
$e$ & 0 (adopted) ($<$\,0.03 at 3$\sigma$) \\ 
$M_{\rm *}$ (M$_{\rm \odot}$) &  1.34 $\pm$ 0.07\\ 
$R_{\rm *}$ (R$_{\rm \odot}$) &  1.29  $\pm$ 0.04\\
$\log g_{*}$ (cgs) & 4.345 $\pm$ 0.019 \\
$\rho_{\rm *}$ ($\rho_{\rm \odot}$) &  0.626 $\pm$ 0.043\\
$T_{\rm eff}$ (K) & 6400 $\pm$ 110\\
$M_{\rm P}$ (M$_{\rm Jup}$) & 0.50 $\pm$ 0.04\\
$R_{\rm P}$ (R$_{\rm Jup}$) &  1.41 $\pm$ 0.05\\
$\log g_{\rm P}$ (cgs) & 2.76 $\pm$ 0.04 \\
$\rho_{\rm P}$ ($\rho_{\rm J}$) & 0.18 $\pm$ 0.02\\
$a$ (AU)  & 0.0506 $\pm$ 0.0009\\
$T_{\rm P, A=0}$ (K) & 1560 $\pm$ 35\\ [0.5mm] \hline 
\multicolumn{2}{l}{Errors are 1$\sigma$; Limb-darkening coefficients were:}\\
\multicolumn{2}{l}{(Trap $z$) a1 =  0.640, a2 = --0.172, a3 = 0.302, 
a4 = --0.174}\\ 
\multicolumn{2}{l}{(Euler $r$) a1 =    0.548, a2 = 0.238, a3 = --0.010, 
a4 = --0.067}\\ \hline
\end{tabular} 
\end{table}

\begin{figure}
\hspace*{-5mm}\includegraphics[width=10cm]{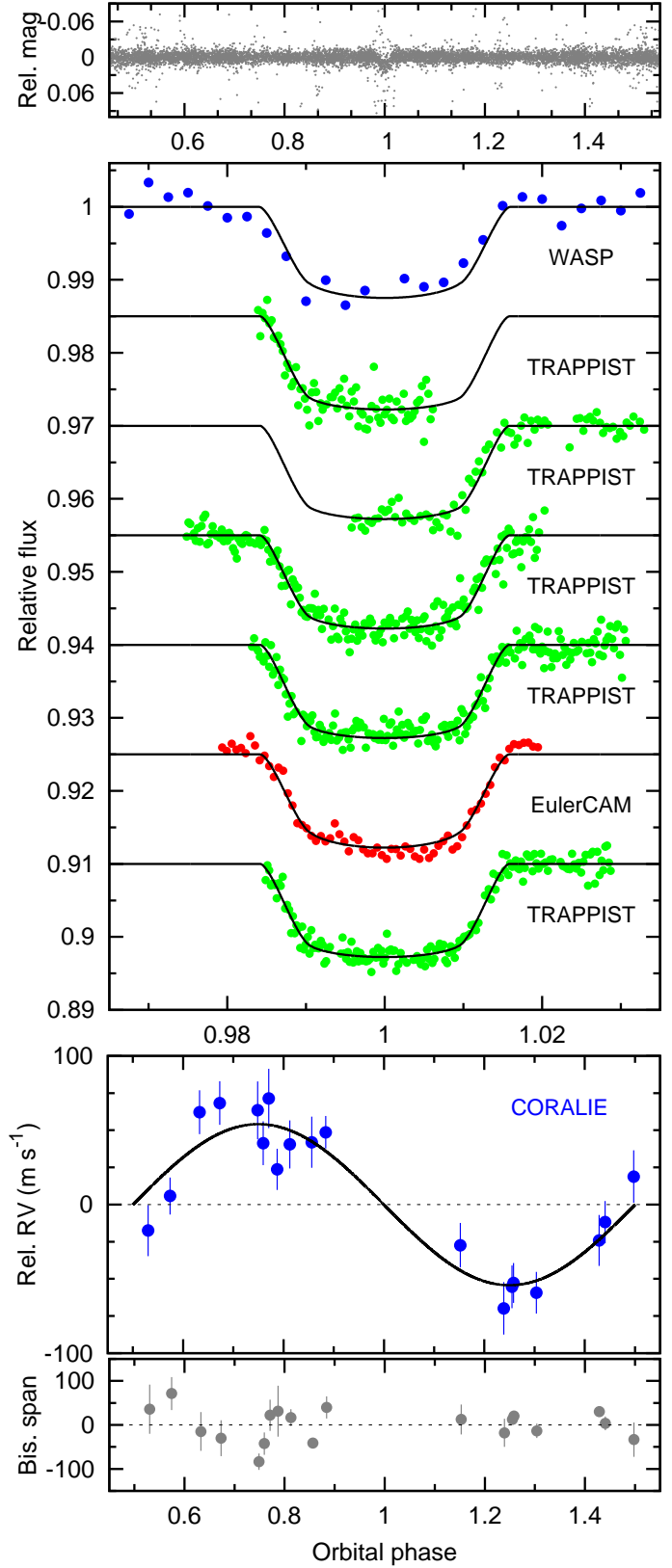}\\ [-2mm]
\caption{WASP-101b discovery data (as in Fig.~1).}
\end{figure}

\clearpage

The fitted parameters were $T_{\rm c}$, $P$, $\Delta F$, $T_{14}$,
$b$, $K_{\rm 1}$, where $T_{\rm c}$ is the epoch of mid-transit, $P$
is the orbital period, $\Delta F$ is the fractional flux-deficit that
would be observed during transit in the absence of limb-darkening,
$T_{14}$ is the total transit duration (from first to fourth contact),
$b$ is the impact parameter of the planet's path across the stellar
disc, and $K_{\rm 1}$ is the stellar reflex velocity
semi-amplitude. 

The transit lightcurves lead directly to stellar
density but one additional constraint is required to obtain stellar
masses and radii, and hence full parametrisation of the system.  Here
we use the calibrations presented by Southworth (2011), based on masses and radii of eclipsing binaries.  

For each system we list the resulting parameters in Tables~2 to 8, and
plot the resulting data and models in Figures~1 to 7.  We also refer
the reader to Smith \etal\ (2012) who present an extensive analysis of
the effect of red noise in the transit lightcurves on the resulting
system parameters.

As in past WASP papers we plot the spectroscopic $T_{\rm eff}$, and
the stellar density from fitting the transit, against the evolutionary
tracks from Girardi \etal\ (2000), as shown in Fig.~8.

%\newpage

%\clearpage

\begin{figure}
\hspace*{-5mm}\includegraphics[width=9.5cm]{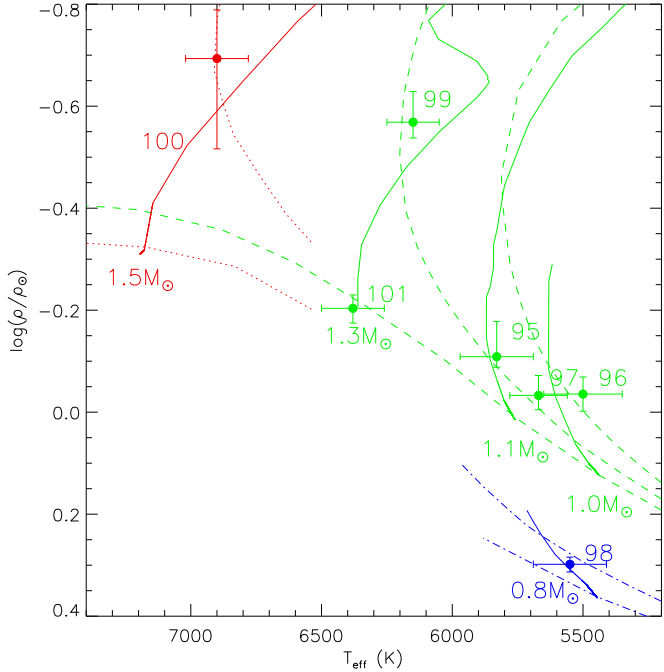}
\caption{Evolutionary tracks on a modified H--R diagram
($\rho_{*}^{-1/3}$ versus $T_{\rm eff}$).  The red lines are for solar 
metallicity, [Fe/H] = 0, showing (solid lines) mass tracks with the labelled mass, and (dashed lines) age tracks for $\log$(age) = 7.85 \&\ 9.15 yrs.
The green lines are the same but for a higher metallicity of [Fe/H] = +0.19 and  $\log$(age) = 7.85, 9.4 \&\ 9.8.   The blue lines are the same for a lower metallicity of [Fe/H] = --0.6, and $\log$(age) = 7.85 \&\ 9.8 yrs. Stars are colour coded to the nearest of these metallicites. The models are from Girardi \etal\ (2000).}
\end{figure}

\subsection{WASP-95}
WASP-95 is a $V$ = 10.1, G2 star with an [Fe/H] of +0.14 $\pm$ 0.16. It may be slightly evolved, with an age of several billion years. 

WASP-95 shows a possible rotational modulation at a period of 20.7 d and an amplitude of 2 mmag in the WASP data (Fig.~9), though this is seen in only one of the two years of data. The values of \vsini\ from the spectroscopic analysis (assuming that the spin axis is perpendicular to us) and the stellar radius from the transit analysis combine to a rotation period of 19.7 $\pm$ 3.9 d, which is compatible with the possible rotational modulation. 

WASP-95b is a typical hot Jupiter ($P_{\rm orb}$ = 2.18 d, $M$ = 1.2 M$_{\rm Jup}$, $R$ = 1.2  R$_{\rm Jup}$). 

\begin{figure}
\hspace*{-5mm}\includegraphics[width=9.5cm]{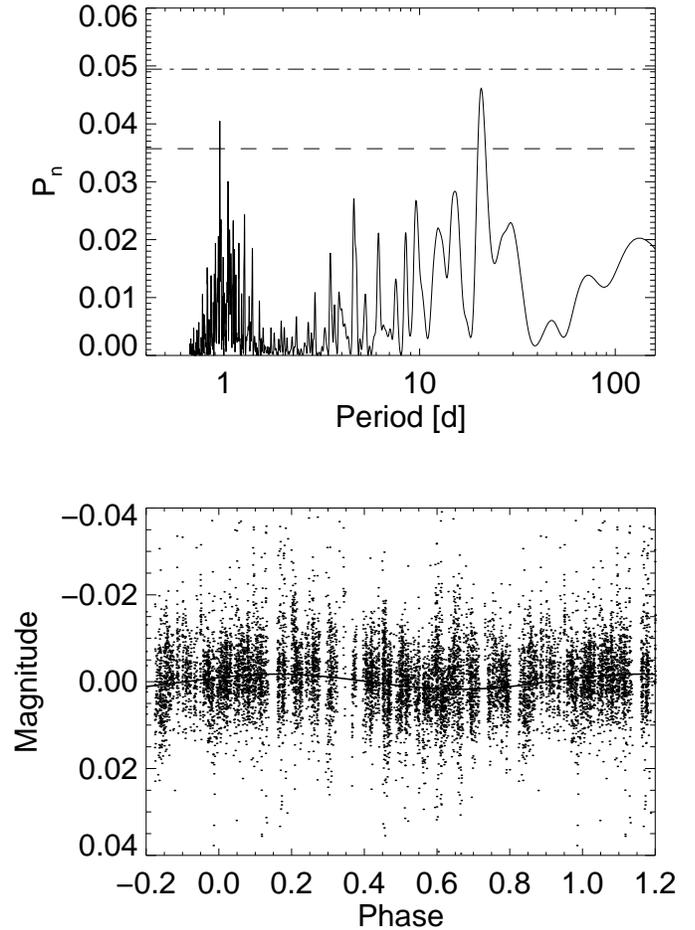}\\ [-2mm]
\caption{The possible rotational modulation in WASP-95 at a period of 20.7 d. The horizontal lines are the 10\%\ (dashed) and 1\%\ (dot-dashed) false-alarm probabilities. The upper plot is the periodogram (the $y$-scale being the fraction of the scatter in the data that is modelled by a sinusoidal variation, weighted by the standard errors); the lower plot is the data folded on the 20.7-d period.}
\end{figure}

\subsection{WASP-96}
WASP-96 is fainter G8 star, at $V$ = 12.2, with  an [Fe/H] of +0.14 $\pm$ 0.19. The planet is a typical hot Jupiter ($P_{\rm orb}$ = 3.4 d, $M$ = 0.5 M$_{\rm Jup}$, $R$ = 1.2  R$_{\rm Jup}$). 

\subsection{WASP-97}
WASP-97 is a  $V$ = 10.6, G5 star, with an above-solar metallicity of [Fe/H] of +0.23 $\pm$ 0.11. The planet is again a typical hot Jupiter ($P_{\rm orb}$ = 2.1 d, $M$ = 1.3 M$_{\rm Jup}$, $R$ = 1.1  R$_{\rm Jup}$). 

\subsection{WASP-98}
At $V$ = 13.0, WASP-98 is at the faint end of the WASP-South/CORALIE survey. The spectral analysis has a lower S/N of only 40 and so is less reliable. It appears to be exceptionally metal poor ([Fe/H] = --0.6 $\pm$ 0.19) for a transit host star.  The spectral analysis suggests G7, though the transit parameters lead to a lower mass of 0.7 M$_{\odot}$.  The planet is a typical hot Jupiter having a high-impact ($b$ = 0.7) transit. 

\subsection{WASP-99}
WASP-99 is the brightest host star reported here, at $V$ = 9.5. Spectral analysis suggests that it is an F8 star with [Fe/H] of +0.21 $\pm$ 0.15, though the transit analysis gives a higher mass ($M$ = 1.5 M$_{\odot}$) and an expanded radius ($R$ = 1.8 R$_{\odot}$).  The planet is relatively massive ($M$ = 2.8 M$_{\rm Jup}$), though its mass and radius are similar to those of many known planets. 

The transit of WASP-99b is the shallowest yet found by WASP-South, at 0.0041 $\pm$ 0.0002, along with that of WASP-72b at 0.0043 $\pm$ 0.0004 (Gillon \etal\ 2013).

\subsection{WASP-100}
WASP-100 is a  $V$ = 10.8, F2 star of solar metallicity.  The planet is a typical bloated hot Jupiter ($P_{\rm orb}$ = 2.8 d, $M$ = 2.0 M$_{\rm Jup}$, $R$ = 1.7  R$_{\rm Jup}$) with a high irradiation (see, e.g., West \etal\ 2013). 

\subsection{WASP-101}
WASP-101 is a  $V$ = 10.3, F6 star with [Fe/H] = +0.20 $\pm$ 0.12. The analysis is compatible with an unevolved main-sequence star.  The planet is a bloated, low-mass planet ($P_{\rm orb}$ = 3.6 d, $M$ = 0.50 M$_{\rm Jup}$, $R$ = 1.4  R$_{\rm Jup}$). 

\section{Completeness}
With the WASP survey now passing WASP-100 we can ask how complete the survey methods are and how many planets we are missing. A full discussion of selection effects is a large topic and is not attempted here, but we can use the HAT project as a straightforward check on our techniques.  The HAT project (Bakos \etal\ 2004) is very similar in conception and hardware to WASP, and thus whether a HAT planet is also detected by us gives an indication of how many we are overlooking. 

\begin{figure}
\hspace*{-5mm}\includegraphics[width=9.2cm]{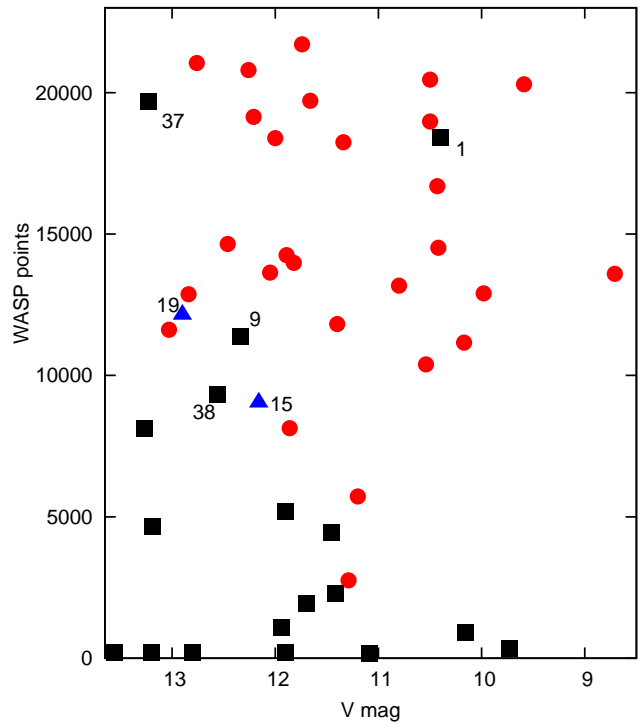}\\ [-2mm]
\caption{The number of WASP data points on each HAT planet as a function of magnitude.  Red circles denote that the planet is detected by WASP, blue triangles are marginal detections, black squares are non-detections.  Some planets are labelled by their HAT number.}
\end{figure}

Fig.~11 shows, for each HAT planet (up to HAT-P-46 and HATS-3b), the number of data points it has in the WASP survey and whether we also detect it.  There is some level of subjectivity in claiming a detection, since all WASP candidates are scrutinised by eye, and thus two planets are shown as marginal detections.  Most non-detections result from there being little or no WASP data ($<$\,6000 data points), but it is worthwhile to review the remaining 7 cases.

HAT-P-1 (Bakos \etal\ 2007) is bright and has 14\,000 points in WASP data. It isn't detected owing to some excess noise in some parts of our automated photometry, which prevents the transit-search algorithm from finding the transits. HAT-P-9b (Shporer \etal\ 2009) has $V$ = 12.3 and there are 11\,000 WASP points.  There is excess noise in some of the WASP photometry, preventing the search algorithm finding the transit.   These two are the worst failures of the WASP techniques applied to the HAT sample. 

HAT-P-37 (Bakos \etal\ 2012) at $V$ = 13.2 is fainter than any WASP planet,  and also has a brighter eclipsing binary 68$^{\prime\prime}$ away, which misleads the WASP search algorithms.  HAT-P-38 (Sato \etal\ 2012) is relatively faint ($V$ = 12.6) and there is relatively limited WASP data (9000 points). 

Of the two marginal detections, HAT-P-19b (Hartman \etal\ 2011) has 12\,000 WASP points but is faint ($V$ = 12.9) and has an orbital period of 4.008 d. As a single-longitude survey, WASP's sampling hampers the finding of integer-day periods.   HAT-P-15b (Kov\'acs \etal\ 2010) is also below 12th magnitude ($V$ = 12.2), and has relatively sparse coverage by WASP (9000 points), and also has a long orbital of 10.86 d, giving fewer transits than the typical 2--5-d hot Jupiters.   

Thus, in summary, at brighter than $V$ = 12.8 and given 10\,000 WASP points (roughly two seasons of data) we fail to detect only 2 out of 23 HAT-discovered planets (note that the number 23 doesn't include independent discoveries by HAT of WASP planets).   Thus we conclude that, in well-covered regions of sky, the WASP-South survey techniques give fairly complete discoveries for the sort of planets findable by WASP-like surveys.  That means hot Jupiters around stars in the magnitude range $V$ = 9--13, and excluding crowded areas of sky such as the galactic plane. It also applies only to stars later than mid F, since hotter stars don't give good RV signals. 

The lower period limit of 0.8 d is likely a real cut-off in the distribution of hot Jupiters (e.g., Hellier \etal\ 2011), since we routinely look for candidates down to periods of 0.5 d.  The longer period limit is primarily set by the amount of observational coverage, and our completeness will drop off above $P_{\rm orb} \sim 7$ d, though this could increase to `warm' Jupiters beyond $\sim$ 10 d as WASP continues to accumulate data.  

The upper limit in planetary size is also likely a real feature of hot Jupiters, given the number of systems found with $R$ = 1.8--2.0 R$_{\rm Jup}$, but not above that range.    The lower size limit of WASP discoveries is set primarily by red noise and so is harder to evaluate.    Fig.~12 shows the transit depths of WASP planets, suggesting that we have a good discovery probability down to depths of $\sim$ 0.5\%\ in the magnitude range $V$ = 9--12.  This is sufficient for  $R_{\rm }$ \sqiggt 0.9 R$_{\rm Jup}$ given $R_{\ast}$ $<$ 1.3 R$_{\sun}$, and  \sqiggt 0.7 R$_{\rm Jup}$ given  $<$ 1.0 R$_{\sun}$, and thus is sufficient for most hot Jupiters.

\begin{figure}
\hspace*{-5mm}\includegraphics[width=9.2cm]{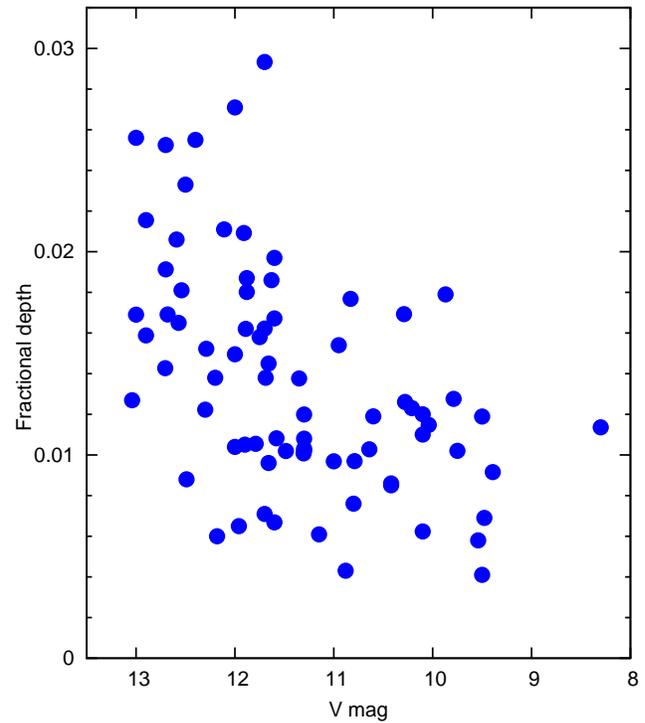}\\ [-2mm]
\caption{The transits depths of WASP planets as a function of host-star magnitude.} \end{figure}

\section*{Acknowledgements}
WASP-South is hosted by the South African
Astronomical Observatory and
we are grateful for their ongoing support and assistance. 
Funding for WASP comes from consortium universities
and from the UK's Science and Technology Facilities Council.
TRAPPIST is funded by the Belgian Fund for Scientific  
Research (Fond National de la Recherche Scientifique, FNRS) under the  
grant FRFC 2.5.594.09.F, with the participation of the Swiss National  
Science Fundation (SNF).  M. Gillon and E. Jehin are FNRS Research  
Associates.  A.H.M.J. Triaud is a Swiss National Science Foundation Fellow under grant PBGEP2-145594

\end{document}